\documentclass[journal,draftcls,onecolumn,12pt,twoside]{IEEEtranTCOM}
\usepackage{booktabs}
\usepackage{multirow}
\usepackage{fancyhdr}
\usepackage{filecontents}
\usepackage[dvipsnames]{xcolor}
\usepackage[percent]{overpic}
\usepackage{tabularx}
\usepackage{arydshln}
\usepackage{verbatim}
\usepackage{gensymb}
\usepackage{stfloats}
\usepackage{mathrsfs}

\normalsize
\usepackage{amsmath,amssymb,amsfonts}
\usepackage{algorithmic}
\usepackage{graphicx, nicefrac}
\usepackage{textcomp}
\usepackage[ruled,vlined]{algorithm2e}
\usepackage{ulem}
\usepackage{setspace} 

\usepackage{color}
\usepackage{array}
\usepackage[version=4]{mhchem}
\usepackage{float}
\usepackage{stmaryrd}
\usepackage{soul}
\usepackage{stackrel}
\usepackage{epstopdf}
\usepackage{cite}
\usepackage{subfig}

\usepackage{bigints}
 
\ifCLASSINFOpdf
\else
\fi

\begin{document}


\title{Heuristic Barycenter Modeling of Fully Absorbing Receivers in Diffusive Molecular Communication Channels}

%
%
%
\author{Fardad~Vakilipoor\IEEEauthorrefmark{1},~\IEEEmembership{Graduate~Student Member,~IEEE},
Abdulhamid~N.M.Ansari, and~
Maurizio~Magarini,~\IEEEmembership{Member,~IEEE}
\thanks{F. Vakilipoor and M. Magarini are with the Department of Electronics, Information and Bioengineering, Politecnico di Milano, I-20133, Milan, Italy e-mail: fardad.vakilipoor@polimi.it, maurizio.magarini@polimi.it.}
\thanks{A. N.M.Ansari is with the Mechanical Engineering Department, University of Hormozgan, Bandar Abbas, Iran e-mail: a.ansari@hormozgan.ac.ir.}
}

\markboth{SUBMITTED TO IEEE TRANSACTIONS ON COMMUNICATIONS}%
{Vakilipoor \MakeLowercase{\emph{et al.}}}

\maketitle

\begin{abstract}
In a recent paper it has been shown that to model a diffusive molecular communication (MC) channel with multiple fully absorbing (FA) receivers, these can be interpreted as sources of negative particles from the other receivers' perspective. The barycenter point is introduced as the best position where to place the negative sources. The barycenter is obtained from the spatial mean of the molecules impinging on the surface of each FA receiver. This paper derives an expression that captures the position of the barycenter in a diffusive MC channel with multiple FA receivers. In this work, an analytical model inspired by Newton's law of gravitation is found to describe the barycenter, and the result is compared with particle-based simulation (PBS) data. Since the barycenter depends on the distance between the transmitter and receiver and the observation time, the condition that the barycenter can be assumed to be at the center of the receiver is discussed. This assumption simplifies further modeling of any diffusive MC system containing multiple FA receivers. The resulting position of the barycenter is used in channel models to calculate the cumulative number of absorbed molecules and it has been verified with PBS data in a variety of scenarios.
\end{abstract}
\IEEEpeerreviewmaketitle
\section{Introduction}
\IEEEPARstart{M}{olecular} Communication (MC) is an interdisciplinary communication paradigm that relies on molecules propagation to exchange information. This unique discipline opens the door to establishing communication on the scale of nanometers to micrometers that can be used between nanorobots or investigating and controlling the natural communications occurring around us. MC studies will lead to reliable cooperation between nano-devices to increase the complexity of their tasks. MC can be divided into two different classes known as natural and artificial. Natural MC has evolved over millions of years to perform various functions in biological systems and there is an excellent potential to investigate it from the communication and information exchange point of view~\cite{bi2021survey}. At the same time, artificial MC is a human-made field that seeks to develop communication systems based on the principles of natural MC. Natural MC occurs in biological systems, where molecules such as hormones, neurotransmitters, and pheromones are used to transmit information between cells, organs, and individuals~\cite{akyildiz2019information}. For example, in the nervous system, neurotransmitters such as dopamine and serotonin are released by neurons to transmit signals to other neurons or to muscle cells~\cite{soldner2020survey}. In the immune system, cytokines and chemokines are released by cells to signal the presence of pathogens or tissue damage~\cite{barros2021molecular}. Artificial MC systems, on the other hand, involve the design and implementation of artificial molecules, often using nanotechnology, to transmit information between artificial entities such as sensors, robots, and implants~\cite{yang2020comprehensive}. In these systems, the artificial molecules are designed to mimic the behavior of natural molecules and interact with artificial receptors to transmit information. One of the advantages of MC is its potential for use in environments where electromagnetic communication is not possible or desirable. For example, MC can be used in applications such as targeted drug delivery, nanomedicine, and implantable devices, where electromagnetic radiation can be harmful or interfere with the operation of the device~\cite{chude2017molecular,cao2019diffusive}. Additionally, molecular communication can be used in underwater environments where electromagnetic waves have limited range and are subject to interference~\cite{guo2020vertical}. MC has been studied under different conditions, e.g. with active or passive receivers, instantaneous or temporal release of molecules, different boundary conditions of the physical channel and so on~\cite{jamali2019channel}.
\subsection{Previous Works}
\par Almost all the works published so far have analyzed Single Input Single Output (SISO) model. However MC systems are intrinsically Multiple Input Multiple Output (MIMO). In fact an MC system with multiple receivers is closer to reality since an individual receiver alone can not afford the complexity of the task. In natural MC that are occurring around us, multiple receivers always cooperate as a unique system. Therefore, it would be a considerable achievement to study multiple receivers and consider their interaction.
\par Modelling the MC with multiple fully absorbing (FA) receivers is a complicated task due to the interaction between the receivers. Some attempts have been made to model the effect of multiple FA receivers in the channel. Authors in~\cite{koo2016molecular} tried to model the channel's impulse response using
a function similar to the SISO system response by applying curve fitting algorithms. Specifically, the paper's channel modeling section used control coefficients over the SISO model to describe MIMO system. The Control coefficients were selected to comprehend system characteristics and being time independent. Then, nonlinear regression models were used to fit simulation data. Similarly~\cite{bao2019channel} relied on the simulation results under different scenarios, and proposed the empirical formulas of the cumulative absorbing probability and the absorption probability density on the intended receiver with respect to the angle between two receivers, distances from transmitter to the receivers, and the spherical receiver size. However, there is no guarantee that such approach can ensure generality of the resulting model. 

Authors in~\cite{sabu2022channel} endeavored to derive an analytical model for a Single Input Multiple Output (SIMO) system operating under conditions of negligibly small mutual interaction between the transmitter and receivers. Such a configuration is tantamount to a scenario wherein the distances separating the receivers from the transceiver, as well as between the receivers themselves, are of a sufficiently great magnitude. However, the key missing part of the generic model for the presence of multiple receivers is that it captures the receivers' observation for any arbitrary positions. They did not consider the cases where receivers are blocking the line of sight of one another or when a receiver is close to the source. Because in that case, the system of the equation describing the model needs to be solved by numerical integration due to strong correlations appearing among the receivers. Moreover, the main limitation of that work and all the related papers available in the literature~\cite{sabu20203,jia2022capacity} is that they miss paying attention to the interaction among the receivers when they are close to the source. To model the diffusive MC system with multiple FA receivers they can be substituted with the negative point sources and their positions are called the barycenters. Previous works did not take into account where to put the negative source. They just put it in the center of the FA receivers (although this false assumption had been discussed and showed thoroughly in the empirical study of~\cite{Fardad} and will be shown with proofs and mathematical discussions in this paper). 

The MC channel model with multiple FA receivers has been recently proposed in~\cite{Fardad}. The authors proposed an analytical model to describe the impulse response of the diffusive channel between a point transmitter and a given number of FA receivers in an MC system. The presence of neighboring FA nanomachines in the environment was taken into account by describing them as point sources of negative molecules. A fundamental problem was the question: ``Where should the negative point source be placed?'' The authors gave an answer to this dilemma by defining the barycenter point, which is the spatial mean of the molecules that have hit the surface of the FA receiver. The authors in~\cite{Fardad} developed an empirical expression to describe the position of the barycenter by applying curve fitting. They found that if there is a source and a receiver in the environment, then the position of the barycenter lies between the center and the surface-point of the spherical receiver. The surface-point is defined as the closest point on the surface of the receiver to the transmitter. They proposed a parameter~$\gamma$ as a function of time and distance between the source and the receiver. The proposed parameter varies between one and zero. When $\gamma$ is equal to one, it means that the spatial average of the molecules from the source are all concentrated at the surface-point of the receiver, while $\gamma$ equal to zero means that the molecules are distributed around the receiver, and their spatial average coincides with the center of the receiver. To obtain an analytical barycenter, one must know the distribution of the particles on the surface of the receiver. 
\subsection{Contributions}
\par In this paper, we derive a heuristic analytical expression that locates the barycenter point of the spherical FA receiver. Having these results allow us to obtain a model that can capture the expected number of absorbed molecules by multiple FA spherical receivers at any arbitrary position. Moreover, the model allows us to have an understanding of under which circumstances we can apply simplifying assumptions and skip the computation of the barycenter and assume it is located at the center of the receivers. This brings the analogy with the common far-field assumption existing in conventional electromagnetic-based communication. First, we describe the system model according to~\cite{Fardad}, then the derivation of the barycenter is shown. Ultimately, we compare the number of molecules absorbed by the receivers based on the resulting model with the analytical barycenter, the empirical barycenter calculated from the particle-based simulation (PBS), and the cumulative number of absorbed molecules obtained directly from it. We considered a single transmitter and two FA receivers MC scenario to ensure that our contribution can be used to describe the presence of a second receiver around the intended receiver. We also compare the analytical $\gamma$ proposed in this paper with the empirical $\gamma$ computed by using the PBS data. Simulation results for the case of two receivers with different radii and five receivers in close proximity are also shown. Hence, despite other available papers in the literature which considered specific scenarios valid for certain conditions, in this paper, we discuss the most complicated scenarios to prove the generality of our model. We believe that using the tools and the methodology introduced in this paper is the missing part to step forward toward modeling the diffusion-based MC with multiple FA receivers.
\subsection{Outline}
\par The rest of the paper is organized as follows. Sec.~\ref{sec:system_model} describes the system model and introduces the negative point source. Then, it starts from SISO modeling and extends it to the scenario with two FA receivers and finally multiple FA receivers. In Sec.~\ref{sec:Barycenter_analysis} we propose the analytical barycenter model, verify it, and investigate its behavior. Sec.~\ref{sec:simulation_results} illustrates the simulation results and validates the model. Finally, Sec.~\ref{sec:conclusion} provides the conclusion of this paper.
\section{System Model}\label{sec:system_model}
In this section we discuss the system model for three cases. To begin with, we review the SISO model. Then, we develop a Single Input Two Output (SITO) model and introduce the concept of negative source. Lastly, we extend the SITO case to a SIMO scenario. The transmitter is point-wise and emits $N_{\mathrm{T}}$ messenger molecules of the same type into the environment instantaneously. The molecules emitted by the transmitter diffuse with constant diffusion coefficient $D~\mu \mathrm{m}^2/\mathrm{s}$ through the medium between transmitter and receivers in an unbounded 3D environment. The receivers have FA properties with a spherical geometry and are able to count the number of absorbed molecules. Once the molecules hit the surface of the receiver, they stop moving. The FA property leads to a coupling effect between the receivers. So to study the number of molecules absorbed by the receivers, we have to take into account the interaction between the receivers.
\subsection{Single Input Single Output (SISO)}
Diffusive molecules propagation is governed by Fick's second law that links the time derivative of the flux to the Laplacian of the molecules' concentration $p\left(r,t\right)$ at distance $r$ and time $t$ as~\cite{redner2001guide}
\begin{equation}
    \frac{\partial p \left( r,t \right)}{\partial t} = D \nabla^2p
    \left( r,t \right). \label{eq:2nd Fick}
\end{equation}
The initial and boundary conditions of~\eqref{eq:2nd Fick} vary depending on the MC system characterization. Yilmaz~\textit{et al.}~\cite{yilmaz2014three} specified the boundary and initial conditions as an impulsive release of molecules, unbounded environment, and an FA spherical receiver $\mathcal{R}$ with radius $R$. They obtained the expression for the hitting rate of the molecules on the surface of the receiver, namely $f\left(d_{(\mathbf{C},\mathbf{\mathcal{T}})},t\right)$, which depends on the distance~$d_{(\mathbf{C},\mathbf{\mathcal{T}})}$ between the transmitter $\mathcal{T}$ and the center of the receiver $\mathcal{R}$, at time~$t$. The channel impulse response of a diffusive MC channel with a single spherical FA receiver of radius $R$ centered at distance $d_{(\mathbf{C},\mathbf{\mathcal{T}})}$ from the transmitter reads
\begin{equation}
f \left( d_{(\mathbf{C},\mathbf{\mathcal{T}})},t \right) = \frac{R \left(d_{(\mathbf{C},\mathbf{\mathcal{T}})}-R\right)}{d_{(\mathbf{C},\mathbf{\mathcal{T}})}\sqrt{4 \pi D
t^3}} e^{-\frac{\left(d_{(\mathbf{C},\mathbf{\mathcal{T}})}-R \right)^2}{4Dt}}, 
\label{eq:imp_SISO}
\end{equation}
and the absorption rate,~\textit{i.e.}, the number of molecules absorbed by the receiver per unit time is
\begin{equation}
n \left( t \right) = N_{\mathrm{T}} f \left( d_{(\mathbf{C},\mathbf{\mathcal{T}})},t \right),
\label{eq:n_R(t)}
\end{equation}
when the transmitter $\mathcal{T}$ emits $N_{\mathrm{T}}$ molecules impulsively. The number of absorbed molecules is obtained by integrating~\eqref{eq:n_R(t)} up to time $t$
 \begin{equation}
 N(t)  =  \int_{0}^t n \left( u \right) \,du =  \frac{N_{\mathrm{T}}R}{d_{(\mathbf{C},\mathbf{\mathcal{T}})}} \mathrm{erfc}\left ( \frac{d_{(\mathbf{C},\mathbf{\mathcal{T}})}-R}{2\sqrt{Dt}} \right )~,\label{eq:integ}
 \end{equation}
where
\begin{equation}
    \mathrm{erfc}\left ( z \right ) \triangleq 1 - \frac{2}{\sqrt{\pi}}\int_{0}^{z} e^{-\tau ^2}d\tau~,
\end{equation}
is the complementary error function.
\subsection{Single Input Two Output (SITO)}
When two receivers are present in the same channel, their absorption rate no longer follows~\eqref{eq:n_R(t)} due to their full absorption characteristic. The presence of a second FA receiver has the effect of removing molecules from the environment, thus reducing the absorption rate of the first receiver.
\par The coupling effect of FA receivers on each other has been taken into consideration by introducing the concept of the negative point source of molecules. To model the negative source effect of receivers, we can consider the existence of a negative point source and replace it with the FA receivers, except for the desired receiver, which must be investigated in terms of the number of absorbed molecules. With reference to the receiver $\mathcal{R}_1$, its hitting rate is influenced both by the number of molecules released by the transmitter and by the reduction of molecules in the environment due to the presence of the receiver $\mathcal{R}_2$. From the $\mathcal{R}_1$ perspective, $\mathcal{R}_2$ can be interpreted as a point source of ``negative'' molecules, characterized by the fact that the number of released molecules coincides with the absorbed ones up to a given time. As shown in~\cite{Fardad}, the best position to place the fictitious negative point source is given by the absorption barycenter point. The barycenter of each receiver is defined as the spatial average of the molecules that adhere to the surface of the receiver due to the FA property. The mutual interaction between the two FA receivers can be modeled by applying the superposition principle.
\par Mathematically, the hitting rate~\eqref{eq:imp_SISO} must be combined with an expression that solves~\eqref{eq:2nd Fick} and satisfies the additional boundary condition at~$\mathcal{R}_2$, which absorbs molecules with an (unknown) absorption rate $n_2\left(t\right)$. The effect of this absorption is accounted for as a negative source. The effect of negative source signal, which is concentrated in the absorption point, perturbs $n_1\left(t\right)$ according to the channel impulse response~\eqref{eq:imp_SISO}. Obviously~\eqref{eq:imp_SISO} is the response to an impulsive release. Since~$n_2\left(t\right)$ varies with time, we take the convolution between them. Because of the symmetry, we can apply the same reasoning by swapping the roles of the absorption rates $n_1\left(t\right)$ and $n_2\left(t\right)$. We can evaluate the absorption rates of the two receivers using
\begin{equation}
    \begin{cases}
        n_1\left(t\right) = N_{\mathrm{T}} f_{(\mathbf{C}_1,\mathbf{\mathcal{T}})}  - n_2\left(t\right) \star f_{(\mathbf{C}_1,\mathbf{B}_2)} 
        \\
        n_2\left(t\right) = N_{\mathrm{T}} f_{(\mathbf{C}_2,\mathbf{\mathcal{T}})}  - n_1\left(t\right) \star f_{(\mathbf{C}_2,\mathbf{B}_1)} 
    \end{cases},
    \label{eq:B_SE}
\end{equation}
where $\star$ denotes the convolution, $f_{(\mathbf{C}_1,\mathbf{\mathcal{T}})}$$\,=\,$$f\left( d_{(\mathbf{C}_1,\mathbf{\mathcal{T}})},t \right)$, $d_{(\mathbf{C}_1,\mathbf{\mathcal{T}})}$ is the distance between the center of $\mathcal{R}_1$ and transmitter, $f_{(\mathbf{C}_1,\mathbf{B}_2)}$$\,=\,$$f\left( d_{(\mathbf{C}_1,\mathbf{B}_2)},t \right)$, and $d_{(\mathbf{C}_1,\mathbf{B}_2)}$ is the distance between the center of $\mathcal{R}_1$ and barycenter of $\mathcal{R}_2$ on the other hand $f_{(\mathbf{C}_2,\mathbf{\mathcal{T}})}$$\,=\,$$f\left( d_{(\mathbf{C}_2,\mathbf{\mathcal{T}})},t \right)$, $d_{(\mathbf{C}_2,\mathbf{\mathcal{T}})}$ is the distance between the center of $\mathcal{R}_2$ and the transmitter, $f_{(\mathbf{C}_2,\mathbf{B}_1)}$$\,=\,$$f\left( d_{(\mathbf{C}_2,\mathbf{B}_1)},t \right)$, and $d_{(\mathbf{C}_2,\mathbf{B}_1)}$ is the distance between the center of $\mathcal{R}_2$ and barycenter of $\mathcal{R}_1$.
\par To determine the expected cumulative number of absorbed molecules on receivers, the integral of~\eqref{eq:B_SE} is required. Taking the Laplace transform of the integration of~\eqref{eq:B_SE}, one obtains
 \begin{equation}
    \begin{cases}
        \hat{N}_1\left(s\right) = \frac{N_{\mathrm{T}} \hat{f}_{(\mathbf{C}_1,\mathbf{\mathcal{T}})}}{s}   - \hat{N}_2\left(s\right)  \hat{f}_{(\mathbf{C}_1,\mathbf{B}_2)}
        \\
        \hat{N}_2\left(s\right) = \frac{N_{\mathrm{T}} \hat{f}_{(\mathbf{C}_2,\mathbf{\mathcal{T}})}}{s}   - \hat{N}_1\left(s\right)  \hat{f}_{(\mathbf{C}_2,\mathbf{B}_1)} 
    \end{cases},
    \label{eq:SITO_S}
\end{equation}
where $\mathscr{L}\{f\}$$\,=\,$$\hat{f}$ and $\mathscr{L}\{N\}$$\,=\,$$\hat{N}$. We can write~\eqref{eq:SITO_S} as a matrix multiplication
\begin{equation}
    \begin{bmatrix}
        \frac{N_{\mathrm{T}}\hat{f}_{(\mathbf{C}_1,\mathbf{\mathcal{T}})}}{s} \\ 
        \frac{N_{\mathrm{T}}\hat{f}_{(\mathbf{C}_2,\mathbf{\mathcal{T}})}}{s}
    \end{bmatrix}
    =
    \begin{bmatrix}
        1 & \hat{f}_{(\mathbf{C}_1,\mathbf{B}_2)}\\ 
        \hat{f}_{(\mathbf{C}_2,\mathbf{B}_1)} & 1 
    \end{bmatrix}
    \begin{bmatrix}
        \hat{N}_{1}\left(s \right )\\ 
        \hat{N}_{2}\left(s \right )
    \end{bmatrix}.\label{eq:1*2Mat}
\end{equation}
Thus the solution in the $S$ domain obtained by a matrix inversion followed by multiplication
\begin{equation}
    \begin{bmatrix}
        \hat{N}_{1}\left(s \right )\\ 
        \hat{N}_{2}\left(s \right )
    \end{bmatrix}
    =
    \begin{bmatrix}
        1 & \hat{f}_{(\mathbf{C}_1,\mathbf{B}_2)}\\ 
        \hat{f}_{(\mathbf{C}_2,\mathbf{B}_1)} & 1 
    \end{bmatrix}^{-1}
    \begin{bmatrix}
        \frac{N_{\mathrm{T}}\hat{f}_{(\mathbf{C}_1,\mathbf{\mathcal{T}})}}{s} \\ 
        \frac{N_{\mathrm{T}}\hat{f}_{(\mathbf{C}_2,\mathbf{\mathcal{T}})}}{s}
    \end{bmatrix}.
    \label{eq:1*2Mat_Inv}
\end{equation}
Applying the inverse Laplace transform on~\eqref{eq:1*2Mat_Inv} results in
    \begin{align}
     \resizebox{0.94\hsize}{!}{$\begin{aligned}
    N_1(t) &= \frac{N_{\mathrm{T}}R_{1}}{d_{(\mathbf{C}_1,\mathbf{\mathcal{T}})}}\sum_{n=0}^{\infty}\left(\frac{R_{1}R_{2}}{d_{(\mathbf{C}_1,\mathbf{B}_2)}d_{(\mathbf{C}_2,\mathbf{B}_1)}}\right)^{n} \mathrm{erfc}\left(\frac{(d_{(\mathbf{C}_1,\mathbf{\mathcal{T}})}-R_{1})+n(d_{(\mathbf{C}_1,\mathbf{B}_2)}+d_{(\mathbf{C}_2,\mathbf{B}_1)}-R_{1}-R_{2})}{2\sqrt{Dt}}\right)\\
    &\hspace{0.4cm}- \frac{N_{\mathrm{T}}R_1R_2}{d_{(\mathbf{C}_1,\mathbf{B}_2)}d_{(\mathbf{C}_2,\mathbf{\mathcal{T}})}}\sum_{n=0}^{\infty}\left(\frac{R_{1}R_{2}}{d_{(\mathbf{C}_1,\mathbf{B}_2)}d_{(\mathbf{C}_2,\mathbf{B}_1)}}\right)^{n}\mathrm{erfc}\left(\frac{(d_{(\mathbf{C}_1,\mathbf{B}_2)}+d_{(\mathbf{C}_2,\mathbf{\mathcal{T}})}-R_{1}-R_{2})+n(d_{(\mathbf{C}_1,\mathbf{B}_2)}+d_{(\mathbf{C}_2,\mathbf{B}_1)}-R_{1}-R_{2})}{2\sqrt{Dt}}\right) .
     \end{aligned}$}\label{eq:SITO_N1_detD}
    \end{align}
that expresses the expected cumulative number of absorbed molecules by $\mathcal{R}_{1}$,
where $R_{1}$ and $R_{2}$ are the radius of receivers $\mathcal{R}_{1}$ and $\mathcal{R}_2$ respectively~\cite[eq. (21)]{Fardad}.
\subsection{Single Input Multiple Output (SIMO)}
Following the same reasoning as for the SITO case, the system of equations corresponding to multiple receivers in the $S$ domain can be written as
    \begin{equation}\label{eq:SIMO_S}
      \left\{\begin{array}{l}
         \hat{N}_1 \left(s\right)   =   \frac{N_{\mathrm{T}} \hat{f}_{(\mathbf{C}_1,\mathbf{\mathcal{T}})}}{s}       -  \hat{N}_2 \left(s\right)   \hat{f}_{{(\mathbf{C}_1,\mathbf{B}_2)}}      - \hat{N}_3 \left(s\right)   \hat{f}_{{(\mathbf{C}_1,\mathbf{B}_3)}} \cdots     -  \hat{N}_p \left(s\right)   \hat{f}_{{(\mathbf{C}_1,\mathbf{B}_p)}}
        \\
         \hat{N}_2 \left(s\right)   =   \frac{N_{\mathrm{T}} \hat{f}_{(\mathbf{C}_2,\mathbf{\mathcal{T}})}}{s}       -  \hat{N}_1 \left(s\right)    \hat{f}_{{(\mathbf{C}_2,\mathbf{B}_1)}}     -  \hat{N}_3 \left(s\right)    \hat{f}_{{(\mathbf{C}_2,\mathbf{B}_3)}} \cdots     -  \hat{N}_p \left(s\right)    \hat{f}_{{(\mathbf{C}_2,\mathbf{B}_p)}}
        \\
        \hspace{1.15cm}\vdots
        \\
         \hat{N}_p \left(s\right)   =   \frac{N_{\mathrm{T}} \hat{f}_{(\mathbf{C}_p,\mathbf{\mathcal{T}})}}{s}       -  \hat{N}_1 \left(s\right)    \hat{f}_{{(\mathbf{C}_p,\mathbf{B}_1)}}     -  \hat{N}_2 \left(s\right)    \hat{f}_{{(\mathbf{C}_p,\mathbf{B}_2)}} \cdots     -  \hat{N}_{p-1} \left(s\right)    \hat{f}_{{(\mathbf{C}_p,\mathbf{B}_{p-1})}},
    \end{array}
    \right.
\end{equation}
Applying the Laplace transform on~\eqref{eq:SIMO_S} allows us to write it in terms of matrix multiplication
\begin{equation}
\resizebox{0.7\hsize}{!}{$
    \begin{bmatrix}
        \hat{N}_{1}\left(s \right )\\ 
        \hat{N}_{2}\left(s \right )\\
        \vdots\\
        \hat{N}_{p}\left(s \right )
    \end{bmatrix}
    =
    \begin{bmatrix}
        1 & \hat{f}_{(\mathbf{C}_1,\mathbf{B}_2)} & \hat{f}_{(\mathbf{C}_1,\mathbf{B}_3)} & \hdots & \hat{f}_{(\mathbf{C}_1,\mathbf{B}_p)}\\ 
        \hat{f}_{(\mathbf{C}_2,\mathbf{B}_1)} & 1 & \hat{f}_{(\mathbf{C}_2,\mathbf{B}_3)} & \hdots & \hat{f}_{(\mathbf{C}_2,\mathbf{B}_p)}\\
        \hat{f}_{(\mathbf{C}_3,\mathbf{B}_1)} & \hat{f}_{(\mathbf{C}_3,\mathbf{B}_2)} & 1 & \hdots & \hat{f}_{(\mathbf{C}_3,\mathbf{B}_p)}\\
        \vdots & \vdots & \vdots & \ddots & \vdots\\
        \hat{f}_{(\mathbf{C}_p,\mathbf{B}_1)} & \hat{f}_{(\mathbf{C}_p,\mathbf{B}_2)} & \hat{f}_{(\mathbf{C}_p,\mathbf{B}_3)} &\hdots & 1
    \end{bmatrix}^{-1}
    \begin{bmatrix}
        \frac{N_{\mathrm{T}}\hat{f}_{(\mathbf{C}_1,\mathbf{\mathcal{T}})}}{s} \\ 
        \frac{N_{\mathrm{T}}\hat{f}_{(\mathbf{C}_2,\mathbf{\mathcal{T}})}}{s} \\
        \vdots\\
        \frac{N_{\mathrm{T}}\hat{f}_{(\mathbf{C}_p,\mathbf{\mathcal{T}})}}{s}
    \end{bmatrix} .
    \label{eq:1*pMat}
    $}
\end{equation}
Unlike the SITO case, the time domain closed-form solution of~\eqref{eq:1*pMat} has not been derived yet. However, since~\eqref{eq:imp_SISO} is causal, we can solve the system of equations numerically.
\section{The Barycenter Analytical Model}\label{sec:Barycenter_analysis}
The main objective of the present paper is to obtain an analytical expression that locates the barycenter of all the receivers an MC system with multiple FA receivers. Because knowing the position of the barycenter allows us to replace other FA receivers with negative point sources and consequently solve~\eqref{eq:SIMO_S}. In the following, we model the barycenter point when there are two FA receivers in the channel by taking advantage of the results from~\cite{saeed2021analytical} and the superposition principle. Then we extend the model to the case of an arbitrary number of FA receivers in the channel.
\subsection{Barycenter in SITO}
\par The barycenter point is the average of the position of molecules that hit the surface of the receiver up to time $t$. It is located inside the receiver's volume and depends on time and the position of the transmitter with respect to both the intended and the other receiver. We define it as the weighted sum of two vectors in 3D space such that the first describes the effect of the transmitter and the second the effect of the other receiver on the intended one. Hence the position of the barycenter point inside the volume of receiver $\mathcal{R}_1$ can be written as
\begin{equation} \label{eq:RR1+RT}
    \mathbf{B}_{1} = \zeta_{(1,1)}\mathbf{B}_{(\mathcal{R}_{1},\mathcal{T})}+ \zeta_{(1,2)}\mathbf{B}_{(\mathcal{R}_{1},\mathcal{R}_{2})} ~,
\end{equation}
where the boldness of a symbol indicates that it represents a vector in 3D space, and $\mathbf{B}_{(\mathcal{R}_{1},\mathbf{\mathcal{T}})}$ corresponds to the barycenter point as the effect of the transmitter on $\mathcal{R}_1$ and $\mathbf{B}_{(\mathcal{R}_{1},\mathcal{R}_{2})}$ corresponds to the $\mathcal{R}_2$'s effect on $\mathcal{R}_1$. Coefficients $\zeta_{(1,1)}$ and $\zeta_{(1,2)}$ must be designed such that represent the contribution of the two sources (\textit{i.e.} the positive source as the result of the transmitter and the negative source as the result of the other receiver) on the barycenter. The positive and the negative source here do not have the same contribution to the position of the barycenter.
\begin{figure}
    \centering
    \includegraphics[width=0.7\columnwidth]{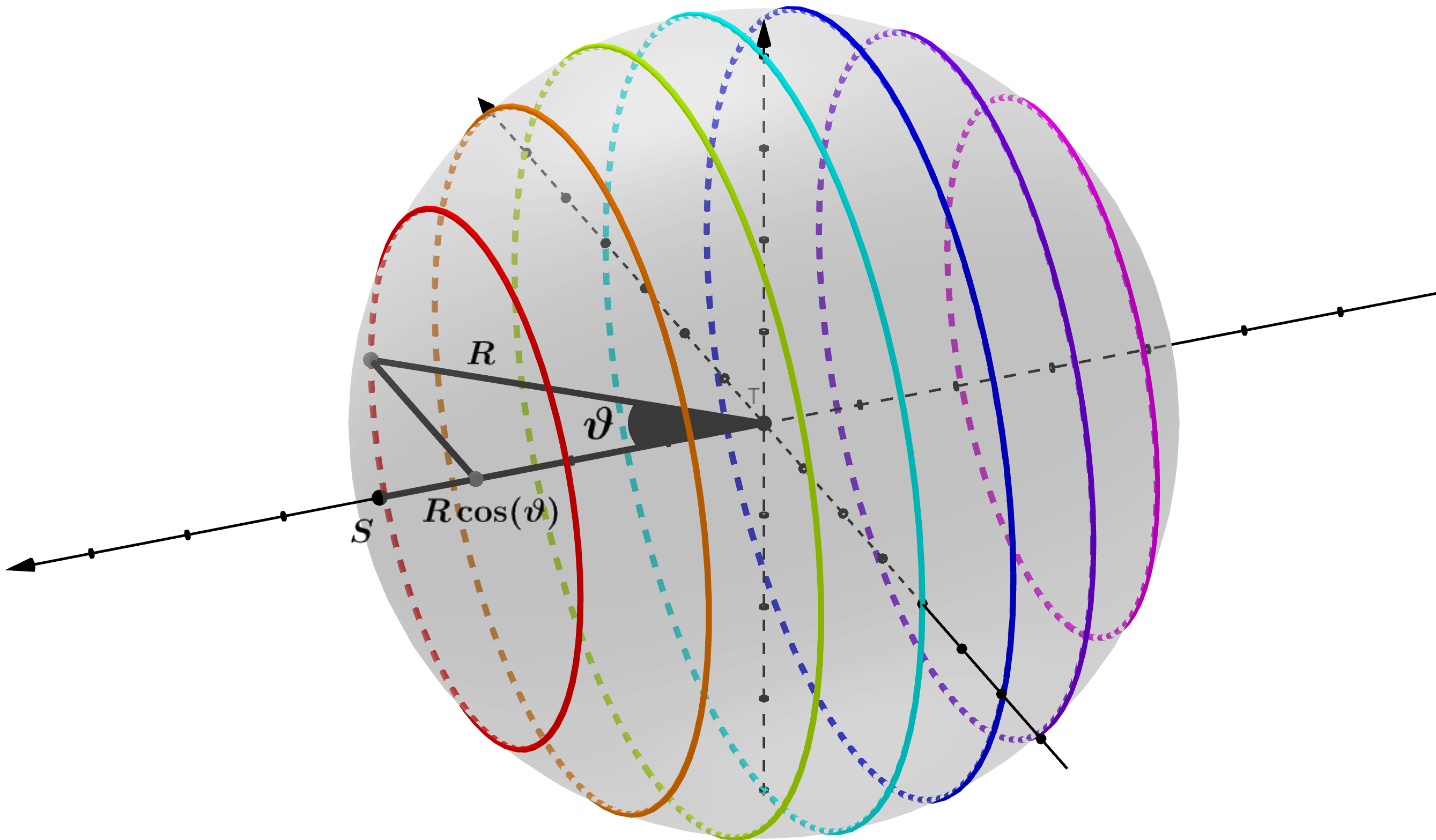}
    \caption{Center of each ring can be formulated as $R\cos{\theta}$ where the angle $\theta$ varies from $0$ to $\pi$.}
    \label{fig:rings}
\end{figure}
\par Let us investigate $\mathbf{B}_{(\mathcal{R}_1,\mathcal{T})}$ assuming that there is no other receiver around. For a spherical receiver, it is clear that $\mathbf{B}_{(\mathcal{R}_1,\mathcal{T})}$ varies on the radius of the sphere towards the transmitter. Intuitively, if we define the closest point on the surface of receiver $\mathcal{R}_1$ to the transmitter $\mathcal{T}$ as the surface-point~$\mathbf{S}_{(\mathcal{R}_1,\mathcal{T})}$ then we can claim that the point $\mathbf{B}_{(\mathcal{R}_1,\mathcal{T})}$ varies from the surface-point $\mathbf{S}_{(\mathcal{R}_1,\mathcal{T})}$, when the transmitter is in touch with the receiver $\mathcal{R}_1$, to the center of the receiver $\mathbf{C}_{1}$, when the distance between the transmitter and receiver is long enough and molecules are spread around the receiver~$\mathcal{R}_1$ uniformly. The variation of the point~$\mathbf{B}_{(\mathcal{R}_1,\mathcal{T})}$ is on the radius of the receiver towards the transmitter due to the physical symmetry of particles propagation in the medium. Thus we write the position of $\mathbf{B}_{(\mathcal{R}_1,\mathcal{T})}$ at time $t$ as
\begin{equation}\label{eq:BT_gamma}
    \mathbf{B}_{(\mathcal{R}_1,\mathcal{T})} = \gamma\left(d_{(\mathbf{C}_1,\mathcal{T})},t\right)\mathbf{S}_{(\mathcal{R}_1,\mathcal{T})} + \left(1-\gamma\left(d_{(\mathbf{C}_1,\mathcal{T})},t\right)\right)\mathbf{C}_{1}~,
\end{equation}
where the parameter $\gamma$ specifies that the point $\mathbf{B}_{(\mathcal{R}_1,\mathcal{T})}$ is located between the surface-point and the center of the spherical receiver $\mathcal{R}_1$. Derivation of the $\gamma$ is explained in the following.
\par To achieve $\gamma$ we need an expression that describes the distribution of the particles over the surface of the receiver. Authors in~\cite{saeed2021analytical} modeled the diffusive MC channel with a single spherical receiver in a spherical coordinate assuming that the receiver is located at the center of the coordinate. However, their goal was not to describe the distribution of the particles over the surface of the receiver. We used their resultant derivation and combined it with our interpretation to have an understanding of the distribution of the particles over the receiver and consequently obtain the $\gamma$. They defined the concentration of the molecules as $p(r,\theta,\phi,t)$ thus, the diffusion equation and its corresponding boundary and initial conditions are
\begin{equation}
    \frac{\partial p \left( r,\theta,\phi,t \right)}{\partial t} = D \nabla^2p
    \left( r,\theta,\phi,t \right). \label{eq:2nd Fick_theta}
\end{equation}
The instantaneous release of the molecules from the point source into the environment at time $t\to 0$ is defined as
\begin{equation}
    p \left( r,\theta,\phi,t\to 0 \right) =     \delta\left( r-d_{(\mathbf{C},\mathbf{\mathcal{T}})} \right)\delta\left( \theta \right)\delta\left( \phi \right). \label{eq:cond_instant}
\end{equation}
The unboundedness of the environment is represented as
\begin{equation}
    \lim_{r\to \infty}p \left( r,\theta,\phi,t\to 0 \right) = 0.\label{eq:cod_unbound}
\end{equation}
The reaction at the surface of the receiver, $r$$\,=\,$$R$, can be described by
\begin{equation}
    D\big( \frac{\partial p\left( r,\theta,\phi,t \right)}{\partial r} + \frac{1}{r}\frac{\partial p\left( r,\theta,\phi,t \right)}{\partial \theta} +  \frac{1}{r\sin{\theta}}\frac{\partial p\left( r,\theta,\phi,t \right)}{\partial \phi}\big)=w p\left( r,\theta,\phi,t \right)~,
    \label{eq:cond_W}
\end{equation}
where $w$ is the reaction rate at the surface of the receiver. To characterize a fully absorbing receiver $w$ should tend to infinite. The solution of the diffusion equation in terms of the cumulative number of absorbed molecules was obtained in spherical coordinate as~\cite[eq.~(33)]{saeed2021analytical}. The solution was written in terms of an integration from~$0$ to~$\pi$. This observation has inspired us to create the surface of a sphere by putting an infinite number of rings as shown in Fig.~\ref{fig:rings} to construct the surface of a spherical receiver.
\par By looking closely at the integration~\cite[eq.~(33)]{saeed2021analytical}, we can decompose the integral based on the angle~$\theta$. Consequently, the fraction of absorbed molecules on rings that create the surface is
\begin{align}
     \resizebox{0.9\hsize}{!}{$\begin{aligned}
Y(\theta,t) = \frac{R^2w}{2(wR+D)}\frac{\alpha+\beta R}{\alpha^{3/2}}\mathrm{erfc}\left(\frac{k}{2\sqrt{t}}\right) &+ \frac{R^2w}{2}\frac{wk^2-\beta(1-mk)}{\alpha Dmk}\\
&\times \exp(mk+m^2t)\mathrm{erfc}\left(m\sqrt{t}+\frac{k}{2\sqrt{t}}\right) ,
    \end{aligned}$} \label{eq:Y}
 \end{align}
where
\begin{equation}
    \alpha = R^2+d^2-2dR\cos{\left(\theta\right)},
\end{equation}
\begin{equation}
    \beta = -R+d\cos{\left(\theta\right)}-d\sin{\left(\theta\right)},
\end{equation}
\begin{equation}
    k = \sqrt{\frac{\alpha}{D}},
\end{equation}
\begin{equation}
    m=\frac{wR+D}{R\sqrt{D}}.
\end{equation}
But~\eqref{eq:Y} depends on $w$ and in order to use it in the case of FA receivers we need to find out $Y$ as $w$$\,\to\,$$\infty$ (See Appendix~\ref{Ap:proof1}).
In the presence of only a transmitter, the spatial average of absorbed particles on the surface of the receiver is expected to be on its radius that is oriented towards the transmitter. This property is due to the symmetrical propagation of particles in all directions around the line of sight. To compute $\gamma$, we consider the absorbed particles of each ring on the surface of the receiver as the weight of the points on the diameter that includes the aforementioned radius. Those points are also the center of each ring. We can take the weighted sum of the coordinate of the circles' centers on the diameter of the sphere aligned with the transmitter as
\begin{equation}
    \gamma\left(d,t\right) = \frac{\bigint_{~0}^{\pi}\left(\frac{\alpha R+\beta R^2}{2\alpha^{3/2}}\mathrm{erfc}\left(\frac{k}{2\sqrt{t}}\right)+ \frac{R^{2}e^{(-\frac{k^2}{4t})}}{2\sqrt{\pi}}\left(\frac{k\sqrt{D}+\beta}{\alpha \sqrt{Dt}}\right)\right)R\cos{\left(\theta\right)}d\theta}{\bigint_{~0}^{\pi}\left(\frac{\alpha R+\beta R^2}{2\alpha^{3/2}}\mathrm{erfc}\left(\frac{k}{2\sqrt{t}}\right)+ \frac{R^{2}e^{(-\frac{k^2}{4t})}}{2\sqrt{\pi}}\left(\frac{k\sqrt{D}+\beta}{\alpha \sqrt{Dt}}\right)\right)Rd\theta} .\label{eq:gamma_td}
\end{equation}
The numerator of~\eqref{eq:gamma_td} is the weighted sum of the points on the diameter of the receiver, which includes the radius of the receiver in the direction of the transmitter. The idea of creating the integral on the numerator is that we want to sum the 1D coordinate of the points~\textit{i.e.,} ranging in $[-R,R]$, on the specified diameter such that each point has a weight that comes from a ring on the surface of the sphere. Note that the aforementioned points are the centers of the rings. To this aim, we take the integral on $\theta$ in the range of~$[0,\pi]$. The relation between the rings and their centers on the specified diameter can be expressed by~$R\cos{\left(\theta\right)}$, which indicates the center of the rings as shown in Fig.~\ref{fig:rings}. Hence we take the integral of all rings as weights of their centers on the diameter. The denominator acts as a normalizing factor because we want that the value of $\gamma$ stays between one and zero. This bound is equivalent to the introduced concept before that the barycenter point in a SISO system remains on the radius between the center ($\gamma=0$) and surface-point ($\gamma=1$) of the receiver.
\par
\begin{figure}
    \centering
    \includegraphics[width=0.6\columnwidth]{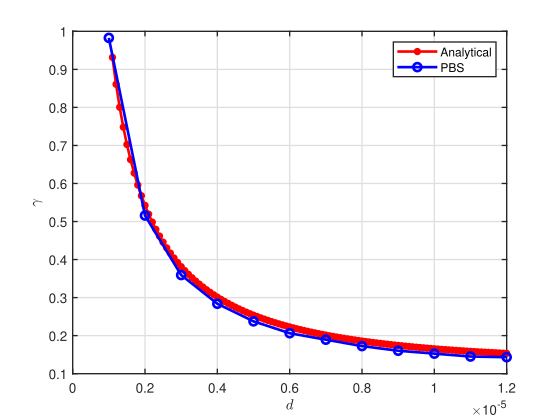}
    \caption{Analytical~$\gamma$ from~\eqref{eq:gamma_td} (the red curve) versus the $\gamma$ obtained from PBS (the blue curve) at $t$$\,=\,$$2$ for different distances between the transmitter and the center of the receiver.}
    \label{fig:gamma_td}
\end{figure}

 Fig.~\ref{fig:gamma_td} compares the empirical $\gamma$ obtained from the PBS with the analytical $\gamma$ from~\eqref{eq:gamma_td}. The PBS data was obtained through a Monte Carlo simulation for~$100$ trials with simulation step time~$10^{-7}\,\mathrm{s}$ and diffusion coefficient~$79.4\,\mu\mathrm{m}^2/\mathrm{s}$. The radius of the receiver is assumed to be~$1\,\mu\mathrm{m}$ and we change the distance between the transmitter and the center of the receiver from~$1$~to $12\,\mu\mathrm{m}$. The duration of PBS is $2\,\mathrm{s}$. As we expected by increasing the distance between the transmitter and the receiver the value of $\gamma$ reduces because the signaling molecules are spread in the space and hit the FA receiver more uniformly compared to the scenario with a close distance between the transmitter and receiver. We would like to note that the slight difference between the empirical $\gamma$ and the analytical one is due to the time discretization of the PBS~\cite{dinc2019effective}. Although we chose simulation step time equal to~$10^{-7}\,\mathrm{s}$ even in this case, not all the particles are trapped over the surface of the FA receiver. Some trap inside the receiver during the PBS. This phenomenon appears due to the discretization of the time domain in PBS while in continuous diffusion we assume particles are absorbed as they hit the surface of the receiver. We would like to underline that the slight observed mismatch is a consequence of temporal discretization in PBS. Finding a solution to deal with the mismatch between Brownian motion and continuous diffusion with absorbing boundary conditions is out of the scope of this paper. 
\par
\begin{figure}
    \centering
    \includegraphics[width=0.6\columnwidth]{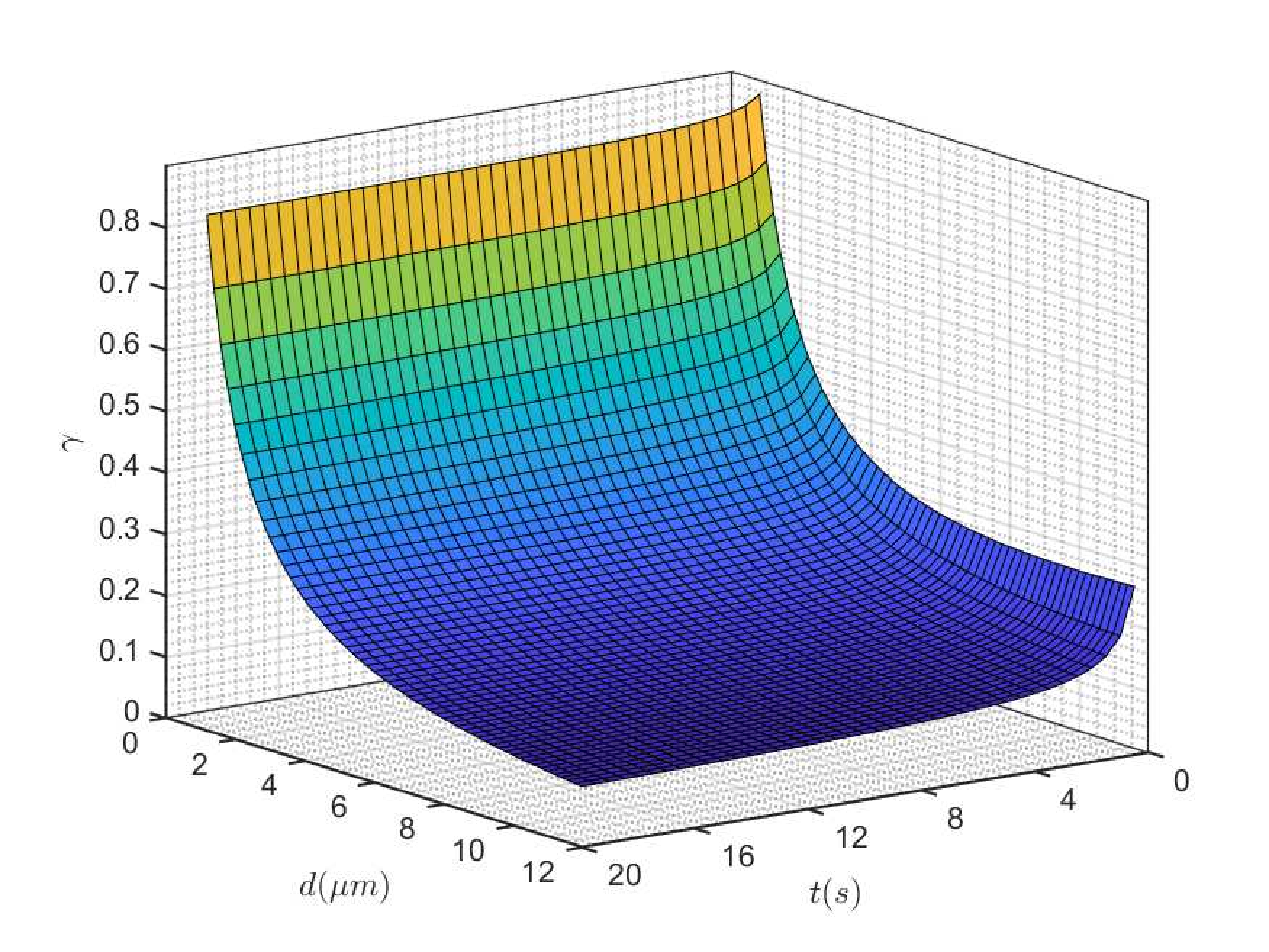}
    \caption{Value of analytical $\gamma$ from~\eqref{eq:gamma_td} as a function of the distance between the source and the center of spherical FA receiver and time.}
    \label{fig:gamma}
\end{figure}

 Fig.~\ref{fig:gamma} shows $\gamma$ for different distances between the source and the receiver and different times of observation. We see that by increasing the time and distance, $\gamma$ tends to zero. As $\gamma$ becomes smaller, according to~\eqref{eq:BT_gamma} the component of the barycenter that represents the effect of the source is approximately located in the center of the receiver.
\par After modeling the effect of transmitter on the receiver, we study the effect of $\mathcal{R}_2$, on~$\mathcal{R}_1$, which was shown by $\mathbf{B}_{(\mathcal{R}_1,\mathcal{R}_2)}$.
We follow a similar approach as the effect of transmitter since we assume that the other receiver performs as a negative point source from the $\mathcal{R}_1$'s perspective. However, in this case instead of having an attraction effect, which was locating the barycenter between the center of $\mathcal{R}_1$ and the surface-point, due to the negativity of the source we assume a repulsion effect. Thus the barycenter point as a result of the negative source is somewhere between the center and the farthest point on the surface of the sphere from the center of~$\mathcal{R}_2$. A simpler way to formulate this definition is just changing the sign of $\gamma$
\begin{equation}
     \mathbf{B}_{(\mathcal{R}_1,\mathcal{R}_2)} = -\gamma\left(d_{(\mathbf{C}_{1},\mathbf{C}_{2})},t\right)\mathbf{S}_{(\mathcal{R}_1,\mathcal{R}_2)} + \left(1+\gamma\left(d_{(\mathbf{C}_{1},\mathbf{C}_{2})},t\right)\right)\mathbf{C}_{1} ,\label{eq:BR_GAMMA}
\end{equation}
where $\mathbf{S}_{(\mathcal{R}_1,\mathcal{R}_2)}$ is the point on the surface of $\mathcal{R}_1$ towards the center of $\mathcal{R}_2$. 
\par We need to combine the contribution of the two sources but it must be taken into account that they do not have the same power to define the position of the barycenter. The effect of the two sources (\textit{i.e.} positive and negative) should not be the same because the number of molecules released by the transmitter is different from that absorbed by the other receiver, which is equivalent to the number of negative molecules. Hence, we consider the coefficients $\zeta_{(1,1)}$ and $\zeta_{(1,2)}$ to equalize the contribution from each source. We propose the value of $1$ to describe the transmitter's effect on the barycenter. On the other hand, to model the effect of the other FA receiver, which is modeled as the fictitious negative source, we have been inspired by Newton's law of universal gravitation~\cite{newton1968mathematical}. Let us consider each receiver as a planet in the universe. Hence the planets can have gravitational forces on each other that is inversely proportional to their squared distance. Moreover another important factor is their mass. But in our case a reasonable parameter that can represent the effective mass of the planets (FA receivers) is their radius. Finally we propose the coefficients to take into consideration the interaction between the receivers as $\nicefrac{R_1 R_2}{d^{2}_{(\mathbf{C}_1,\mathbf{C}_2})}$. By normalizing the coefficients we have
\begin{equation}
    \zeta_{(1,1)} = \frac{1}{1+\frac{R_1 R_2}{d^{2}_{(\mathbf{C}_1,\mathbf{C}_2)}}}~,
\end{equation}
\begin{equation}
    \zeta_{(1,2)} = \frac{\frac{R_2 R_1}{d^{2}_{(\mathbf{C}_1,\mathbf{C}_2})}}{1+\frac{R_1 R_2}{d^{2}_{(\mathbf{C}_1,\mathbf{C}_2)}}}~.
\end{equation}
The barycenters of receivers in a SITO system becomes
\begin{equation} \label{eq:RR1+RT_SITO}
    \mathbf{B}_{1} = \frac{\mathbf{B}_{(\mathcal{R}_1,\mathcal{T})}+ \frac{R_1 R_2}{d^{2}_{(\mathbf{C}_1,\mathbf{C}_2)}}\mathbf{B}_{(\mathcal{R}_1,\mathcal{R}_2)}}{1+\frac{R_1 R_2}{d^{2}_{(\mathbf{C}_1,\mathbf{C}_2)}}} ~,
\end{equation}
\begin{equation} \label{eq:RR2+RT_SITO}
    \mathbf{B}_{2} = \frac{\frac{R_2 R_1}{d^{2}_{(\mathbf{C}_2,\mathbf{C}_1)}}\mathbf{B}_{(\mathcal{R}_2,\mathcal{R}_1)} + \mathbf{B}_{(\mathcal{R}_2,\mathcal{T})} }{\frac{R_2 R_1}{d^{2}_{(\mathbf{C}_2,\mathbf{C}_1)}}+1} ~.
\end{equation}
\par By looking at~\eqref{eq:BT_gamma} and~\eqref{eq:BR_GAMMA} we understand that the barycenter is highly dependent on the behavior of $\gamma$ contributed by the transmitter and the other receiver. According to Fig.~\ref{fig:gamma} we observe that by increasing the distance and time the value of $\gamma$ gets closer to zero. Hence, if the distance between the receivers themselves and the transmitter is not extremely close after a certain time of observation we can assume that the barycenter is located at the center of the spherical FA receivers. In fact, authors in~\cite{Fardad_Asym} investigated an asymptotic model for the MIMO MC system and assumed that the barycenter position could be approximated as the center of the receivers. Their justification was based on the definition of the concept and the empirical observations. From Fig.~\ref{fig:gamma} we conclude that if the distance between the source is not very close, then by increasing the observation time we can assume that the barycenter is located at the center of the receiver. Studying $\gamma$ and its variation contributed from the transmitter and the other receivers allow to simplify the modeling process and even skip the computation of the barycenter under certain conditions such as temporal asymptotic models~\cite{Fardad_Asym}. This analysis is one of the main contributions of this paper that allows researchers to have a tool in order to simplify their computation while investigating diffusive MC with multiple spherical FA receivers.
\subsection{Barycenter in SIMO}
In case of $p$ FA receivers in the channel, we can write the position of the barycenter $\mathbf{B}_{i}$ of $\mathcal{R}_i$, as the weighted sum of contributions from the transmitter and other receivers (similar to~\eqref{eq:RR1+RT})
\begin{equation} \label{eq:RRi+RT}
     \mathbf{B}_{i} = \zeta_{(i,i)}\mathbf{B}_{(\mathcal{R}_{i},\mathcal{T})}+ \sum_{\substack{j=1 \\ j\neq i}}^{p} \zeta_{(i,j)}\mathbf{B}_{(\mathcal{R}_{i},\mathcal{R}_{j})}~.
\end{equation}
The effect of the transmitter on the barycenter of $\mathcal{R}_i$ is 
\begin{equation}\label{eq:BTi_gamma}
    \mathbf{B}_{(\mathcal{R}_i,\mathcal{T})} = \gamma\left(d_{(\mathbf{C}_i,\mathcal{T})},t\right)\mathbf{S}_{(\mathcal{R}_i,\mathcal{T})} + \left(1-\gamma\left(d_{(\mathbf{C}_i,\mathcal{T})},t\right)\right)\mathbf{C}_{i}~.
\end{equation}
The effect of~$\mathcal{R}_j$ on $\mathcal{R}_i$ is 
\begin{equation}
     \mathbf{B}_{(\mathcal{R}_i,\mathcal{R}_j)} = -\gamma\left(d_{(\mathbf{C}_{i},\mathbf{C}_{j})},t\right)\mathbf{S}_{(\mathcal{R}_i,\mathcal{R}_j)} + \left(1+\gamma\left(d_{(\mathbf{C}_{i},\mathbf{C}_{j})},t\right)\right)\mathbf{C}_{i}.\label{eq:BRij_GAMMA}
\end{equation}
Accordingly the coefficients $\zeta_{(i,i)}$ and $\zeta_{(i,j)}$ are
\begin{equation}
    \zeta_{(i,i)} = \frac{1}{1+\Sigma_{k=1}^{p}\frac{R_i R_k}{d^2_{(\mathbf{C}_{i},\mathbf{C}_{k})}}} ,~~k\neq i
\end{equation}
and
\begin{equation}
    \zeta_{(i,j)} = \frac{\frac{R_i R_j}{d^2_{(C_i,C_j)}}}{1+\Sigma_{k=1}^{p}\frac{R_i R_k}{d^2_{(\mathbf{C}_{i},\mathbf{C}_{k})}}}.~~k\neq i
\end{equation}
\begin{figure}
    \centering
    \includegraphics[width=0.6\columnwidth]{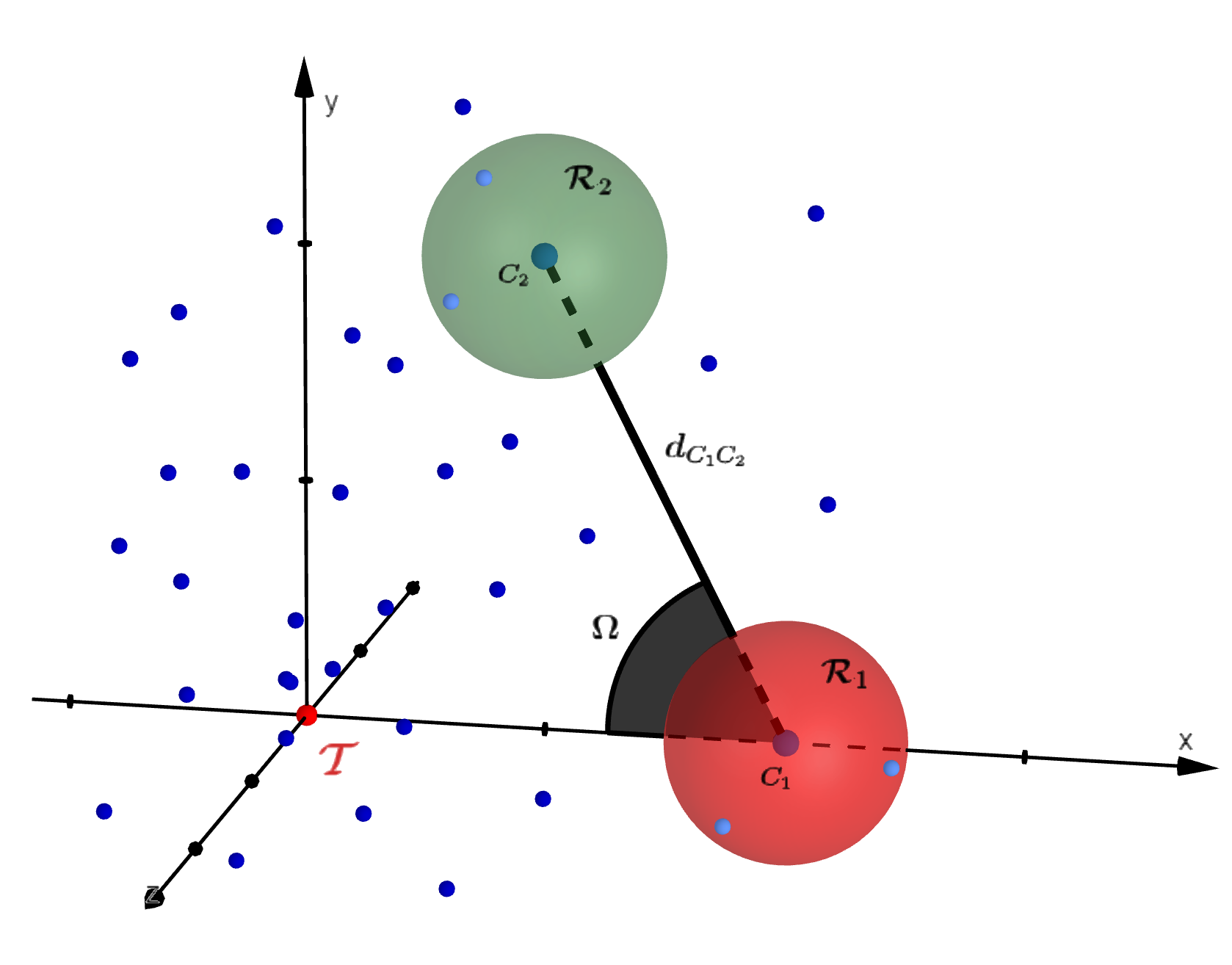}
    \caption{A diffusive MC system with two FA receivers centered at points $\mathbf{C}_1$ and $\mathbf{C}_2$.}
    \label{fig:topology}
\end{figure}
\section{Simulation and Results}\label{sec:simulation_results}
\begin{figure}[t!]
    \centering
    \includegraphics[width=1\columnwidth]{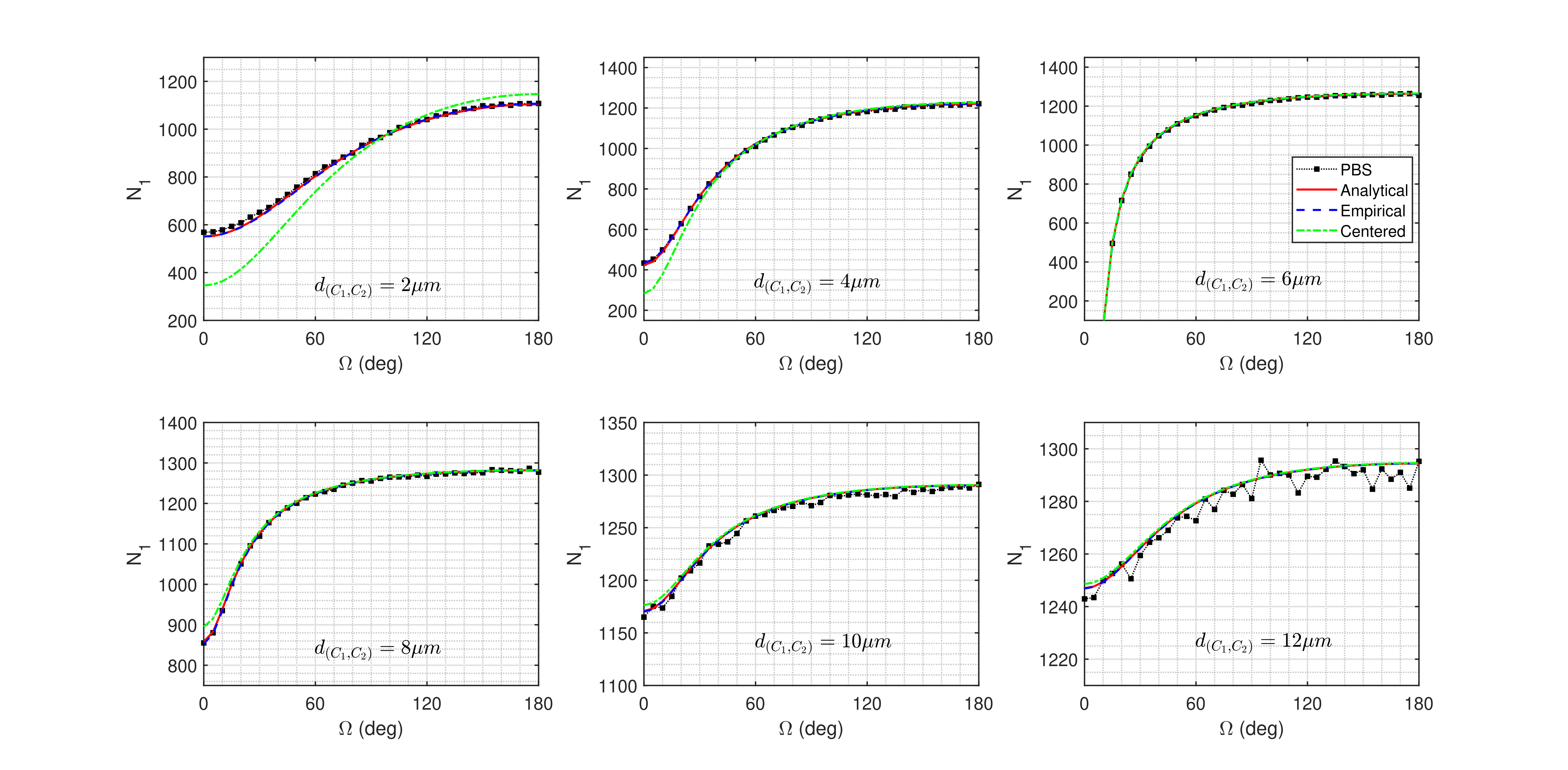}
    \caption{Cumulative expected number of molecules $N_1$ absorbed by $\mathcal{R}_1$ after $t$$\,=\,$$2\,\mathrm{s}$, according to the scenario of Fig.~\ref{fig:topology} with $d_{(\mathbf{C}_1,\mathcal{T})}$$\,=\,$$6\,\mu\mathrm{m}$, and $R_1$$\,=\,$$R_2$$\,=\,$$1\,\mu\mathrm{m}$ for various positions of $\mathcal{R}_2$ identified by the distance $d_{(\mathbf{C}_1,\mathbf{C}_2)}$$\,=\,$$\{2,4,6,8,10,12\}\,\mu\mathrm{m}$ and angle $\Omega$$\,=\,$$[0,180]\degree$. The red line, ``Analytical'', shows the cumulative expected number of absorbed molecules by $\mathcal{R}_1$ based on~\eqref{eq:SITO_N1_detD} by calculating the barycenters positions analytically. The blue line, ``Empirical'', is using the same equation as the red line but the position of the barycenters are calculated empirically from the distribution of the particles over the receivers during the PBS. The dotted line with black squares, ``PBS'', is the cumulative number of absorbed molecules by $\mathcal{R}_1$. The green dash-dot line, ``Centered'', is the value of $N_1$ when we assume the barycenter is located at the center of the receiver.}
    \label{fig:Same_R}
\end{figure}
\par In this section we verify the proposed model with the data obtained from PBS. In order to show the advantage of knowing the barycenter we compute the case where most of the available works assumed that the negative source is simply in the centers of the receivers. Some of the simulation parameters are given in Tab.~\ref{tab:param}, and they are borrowed from~\cite{Fardad}, except for PBS step time that we reduced even by one more order of magnitude compared to the one used in~\cite{Fardad} to have better precision in the simulation results. We demonstrate the position of the barycenters obtained from the PBS in different scenarios and compare them with the analytical model. Moreover, we compare the expected cumulative number of absorbed molecules by the receivers with the data obtained from PBS. All the results from PBS are averaged over $100$ trials and the simulation step time is $10^{-7}\,\mathrm{s}$. The transmitter is located at the center of the coordinate. In order to simplify the visualizations we assume that all the receivers are located in $xy$\nobreakdash-plane. However, the channel is 3D and the molecules move in all directions.
Fig.~\ref{fig:topology} depicts the simulation scenario of SITO system. The center of $\mathcal{R}_1$ denoted by $\mathbf{C}_1$ is fixed on $X$ axis at coordinate $\left(d_{(\mathbf{C}_1,\mathcal{T})},0,0\right)$.
\par The expected cumulative number of absorbed molecules by $\mathcal{R}_1$ at time $t$$\,=\,$$2\,\mathrm{s}$ for different~$\Omega$ and~$d_{(\mathbf{C}_1,\mathbf{C}_2)}$ is depicted in Fig.~\ref{fig:Same_R}. The horizontal axis of each subfigure represents the angle $\Omega$ from $0\degree$ to $180\degree$, and the vertical axis is the cumulative number of absorbed molecules by the receiver $\mathcal{R}_1$. The radii of the receivers are the same and equal to $1\,\mu\mathrm{m}$. Each subfigure corresponds to the different~$d_{(\mathbf{C}_1,\mathbf{C}_2)}$. The black dotted line with square marker represents the average value of the cumulative number of absorbed molecules by $\mathcal{R}_1$ that we obtained from PBS. The empirical barycenter is then substituted onto~\eqref{eq:SITO_N1_detD} and the blue line shows the expected cumulative number of absorbed molecules based on the empirical barycenter that was obtained through the PBS. The red line is drawn based on~\eqref{eq:SITO_N1_detD} while the barycenter is computed analytically as the main contribution of this paper. The green dash-dot line is when the barycenter position is substituted with the center of the receivers. We can observe an accurate match between the PBS and the analytical results. The variation of the PBS data in case of large distances is just due to the stochasticity of the Brownian motion and considering the range of variation it can be neglected. It can be seen that in case of the line of sight blockage of $\mathcal{R}_1$ by $\mathcal{R}_2$ the distribution of the particles changes severely and consequently the assumption of considering barycenter at the center of the sphere is no more valid. This observation confirms our discussion in Sec.~\ref{sec:Barycenter_analysis} about the behavior of the $\gamma$ when the distribution of the absorbed particles over the surface of the receiver is not uniform. The non-uniformity of the particles can be due to the closeness of a source to the receiver or the time of observation.
\begin{table}
\begin{center}
\vspace{.1cm}
\caption{Simulation parameters}
\vspace{-.1cm}
\label{tab:param}
\resizebox{0.6\textwidth}{!}{
 \begin{tabular}{|| c | c | c ||}
 \hline
 Variable & Definition & Value \\ [0.5ex] 
 \hline\hline
  $N_{\mathrm{T}}$ & Number of released molecules & $10^{4}$ \\ 
  \hline
 $R$ & Receivers radius & $1\,\mu\mathrm{m}$\\ 
 \hline
 $D$ & Diffusion coefficient for the signaling molecule & $79.4\,\mu \mathrm{m}^2/\mathrm{s}$ \\
 \hline
 $\Delta t$ & PBS step time & $0.1\,\mu \mathrm{s}$\\
 \hline
\end{tabular}}
\end{center}
\vspace{-.2cm}
\end{table}
\begin{figure}[t]
    \centering
    \includegraphics[width=1\textwidth]{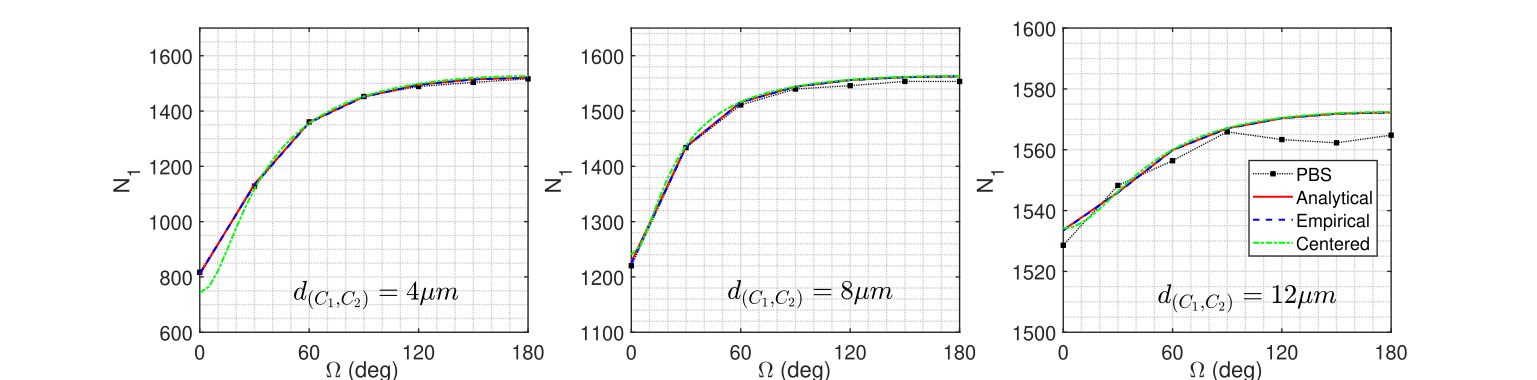}
    \caption{Cumulative expected number of molecules $N_1$ absorbed by $\mathcal{R}_1$ after $t$$\,=\,$$2\,\mathrm{s}$, according to the scenario of Fig.~\ref{fig:topology} with $d_{(\mathbf{C}_1,\mathcal{T})}$$\,=\,$$6\,\mu\mathrm{m}$, $R_1$$\,=\,$$1.2\,\mu\mathrm{m}$, and $R_2$$\,=\,$$0.7\,\mu\mathrm{m}$ for various positions of $\mathcal{R}_2$ identified by the distance $d_{(\mathbf{C}_1,\mathbf{C}_2)}$$\,=\,$$\{4,8,12\}\,\mu\mathrm{m}$ and angle $\Omega$$\,=\,$$[0,180]\degree$. The legend of the figures is defined the same as the ones explained in Fig.~\ref{fig:Same_R}.}
    \label{fig:different_R}
\end{figure}
\par In Fig.~\ref{fig:different_R} we depict a similar output as Fig.~\ref{fig:Same_R} when the radius of the receivers are not the same. In this case we assumed that the radius of the receiver $\mathcal{R}_1$ is $1.2\,\mu\mathrm{m}$ and the radius of the receiver $\mathcal{R}_2$ is $0.7\,\mu\mathrm{m}$. Even in this case we observe a good match between the analytical and PBS results. The legend and markers are the same as the ones described in Fig.~\ref{fig:Same_R}. Note that the slight deviation of the PBS data in the last subfigure corresponding to $d_{(\mathbf{C}_1,\mathbf{C}_2)}$$\,=\,$$12\,\mu\mathrm{m}$ is due to the stochasticity of Brownian motion. Moreover, the range of variation is minor. When $d_{\mathbf{C}_1,\mathbf{C}_2}$$\,=\,$$4\,\mu\mathrm{m}$ and $\Omega$ is between $0\degree$ to $30\degree$ we can observe the difference between the ``Centered'' line and the PBS data. It is exactly when $\mathcal{R}_2$ is blocking the line of sight between the source and $\mathcal{R}_1$.
\par
\begin{figure}
    \centering
    \includegraphics[width=0.7\columnwidth]{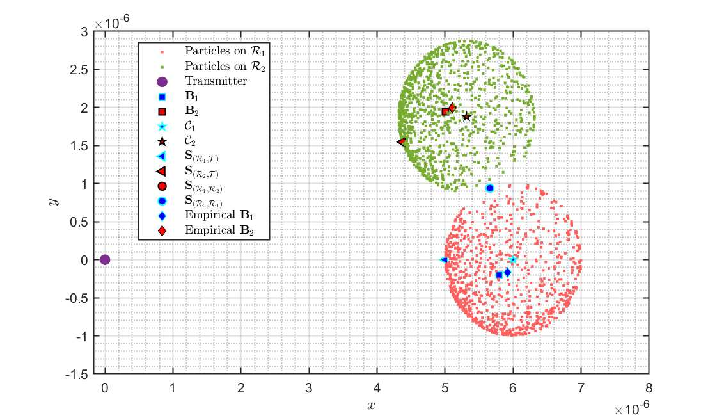}
    \caption{Empirical and analytical barycenters when $d_{(\mathbf{C}_1,\mathbf{C}_2)}$$\,=\,$$2\,\mu\mathrm{m}$, $\Omega$$\,=\,$$70\degree$, and $t$$\,=\,$$2\,\mathrm{s}$. Small dots indicate the distribution of the absorbed particles over the surface of the receivers.}
    \label{fig:bary_d2_th70}
\end{figure}

 In Fig.~\ref{fig:bary_d2_th70} we show the position of the empirical barycenter based on the distribution of the particles on the surface of the receivers obtained during the PBS and also depict the analytical barycenter position. In this scenario the two receivers have the radius of $1\,\mu\mathrm{m}$, and they are in touch with each other meaning that $d_{\mathbf{C}_1,\mathbf{C}_2}$$\,=\,$$2\,\mu\mathrm{m}$. The angle $\Omega$ is $70\degree$. The reason we decided to show the details of this scenario was that based on our observations when the receivers are in touch, it becomes the most difficult case to model and obtain the results with tolerable accuracy. We observe that even in this critical case the analytical barycenters (squared markers) are very close to the empirical ones (diamond markers).
\begin{figure}[t] 
    \centering
  \subfloat[\label{1a}]{%
       \includegraphics[width=0.45\columnwidth]{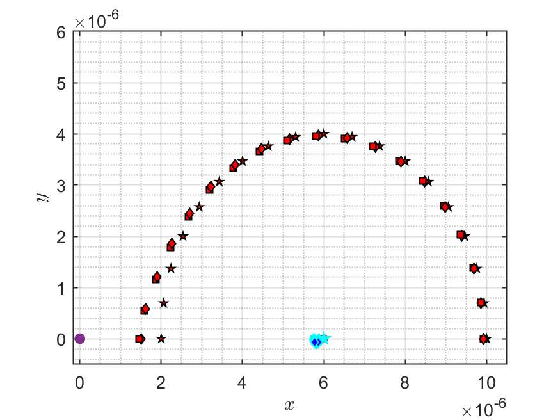}}
  \subfloat[\label{1b}]{%
        \includegraphics[width=0.45\columnwidth]{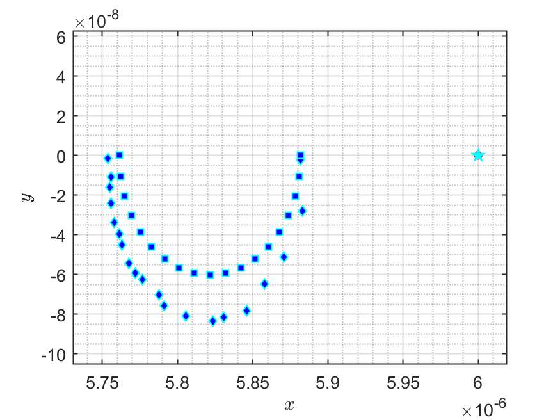}}
  \caption{(a) Positions of the analytical (squares) and empirical (diamond) barycenters, $\mathbf{B}_2$, corresponding to different positions of the $\mathcal{R}_2$. The pentagram indicates the center of $\mathcal{R}_2$. Receiver $\mathcal{R}_2$ revolves around $\mathcal{R}_1$ at distance $d_{C_1 C_2}$$\,=\,$$4$. (b) Magnified position of the empirical (diamond) and estimated (square) barycenters, $\mathbf{B}_1$.}\label{fig:B_R_d4}
\end{figure}
\par In Fig.~\ref{fig:B_R_d4}a we demonstrate the variation of both analytical and empirical barycenters by changing the position of $\mathcal{R}_2$. In this scenario $d_{(\mathbf{C}_1,\mathbf{C}_2)}$$\,=\,$$4\,\mu\mathrm{m}$ and $\Omega$ varies from $0\degree$ to $180\degree$. The legend of the figure is the same as the one in~Fig.~\ref{fig:bary_d2_th70}. In Fig.~\ref{fig:B_R_d4}a the red stars are the center of~$\mathcal{R}_2$, while the red squares are the analytical barycenters, and the red diamonds are the empirical barycenters. We can observe that when $\Omega$$\,=\,$$0\degree$ the attraction effect from the transmitter is strong since the distance between the receiver and transmitter is not that long. We can see that as $\mathcal{R}_2$ moves behind the $\mathcal{R}_1$ (the cyan star) with respect to the transmitter (the purple circle), the barycenter moves towards the center of the receiver $\mathcal{R}_2$. When $\Omega$$\,=\,$$180\degree$ we see that the barycenter converges to the center of the receiver since the distance between $\mathcal{R}_2$ and the transmitter increases. The repulsion effect of $\mathcal{R}_1$ on the barycenter of $\mathcal{R}_2$ is negligible in this case compared to that of the transmitter's effect. 
\par Fig.~\ref{fig:B_R_d4}b shows the variation of the barycenter $\mathbf{B}_1$ corresponding to different positions of $\mathcal{R}_2$ shown in Fig.~\ref{fig:B_R_d4}a. As the position of the $\mathcal{R}_2$ changes clockwise the position of the barycenter $\mathbf{B}_1$ also changes clockwise. Note that Fig.~\ref{fig:B_R_d4}b is a zoomed version of a small area inside $\mathcal{R}_1$ and the difference between the analytical barycenters (the blue squares) and the empirical ones (the blue diamonds) are maximum $3\cdot10^{-2}\,\mu\mathrm{m}$. Moreover, we can notice from both Fig.~\ref{fig:B_R_d4}a and \ref{fig:B_R_d4}b that our proposed model of analytical barycenter captures the variation corresponding to the position of the receivers in the channel.
\par
\begin{figure}
    \centering
    \includegraphics[width=0.7\columnwidth]{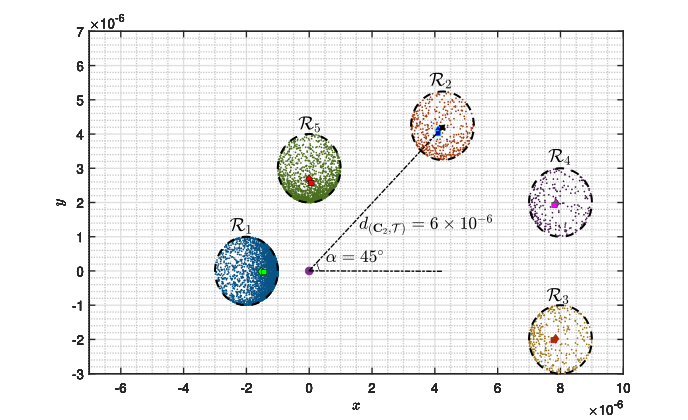}
    \caption{Topology of simulation scenario with five receivers and one transmitter. The transmitter is at the center of the coordinate. Receivers $\mathcal{R}_i$, $i$$\,\in\,$$\{1,3,4,5\}$ are at $(-2\times10^{-6},0)$, $(8\times10^{-6},-2\times10^{-6})$, $(8\times10^{-6},2\times10^{-6})$, and $(0,3\times10^{-6})$ in $xy$\nobreakdash-plane respectively. Position of $\mathcal{R}_2$ varies for different angles of $\alpha$ in each scenario (in this figure $\alpha$$\,=\,$$45\degree$). The distance between the center of $\mathcal{R}_2$ and the transmitter is equal to $6\,\mu\mathrm{m}$. Squared markers indicate the analytical barycenter point and diamond markers indicate the empirical barycenter obtained by taking the average of the position of particles that are absorbed by the corresponding receivers.}
    \label{fig:topo_si50}
\end{figure}

 To verify the model for a scenario with high complexity we considered a simulation topology with five receivers in the environment. Fig.~\ref{fig:topo_si50} depicts the new simulation scenario where four of the receivers are fixed and receiver $\mathcal{R}_2$ changes its position based on the angle $\alpha$ in every simulation trial. The parameter $\alpha$ varies from $0\degree$ to $180\degree$ and the distance between the center of $\mathcal{R}_2$ and the transmitter, which is located at the center of the coordinate is fixed to $6\,\mu\mathrm{m}$. The distributions of absorbed particles are shown for the specific example depicted in the figure at $t$$\,=\,$$2\,\mathrm{s}$. We can observe a very good estimation of the analytical barycenters (the squared markers) and the empirical barycenters (the diamond markers). We deliberately designed this configuration of receivers because we found that in the extreme cases when receivers are very close to the source or one another the estimation of the barycenter and consequently the cumulative number of absorbed molecules becomes difficult. Hence, we put $\mathcal{R}_1$ very close to the source. Moreover we considered the position of $\mathcal{R}_2$ such that for small angles of $\alpha$ it blocks the line of sight of $\mathcal{R}_4$ and for the high value of $\alpha$ it goes to the shadowing area of $\mathcal{R}_1$ and $\mathcal{R}_5$. Hence we believe that the proposed scenario is a good test to challenge different aspects of the proposed model. 
\par
\begin{figure}[ht!]
    \centering
    \includegraphics[width=1\columnwidth]{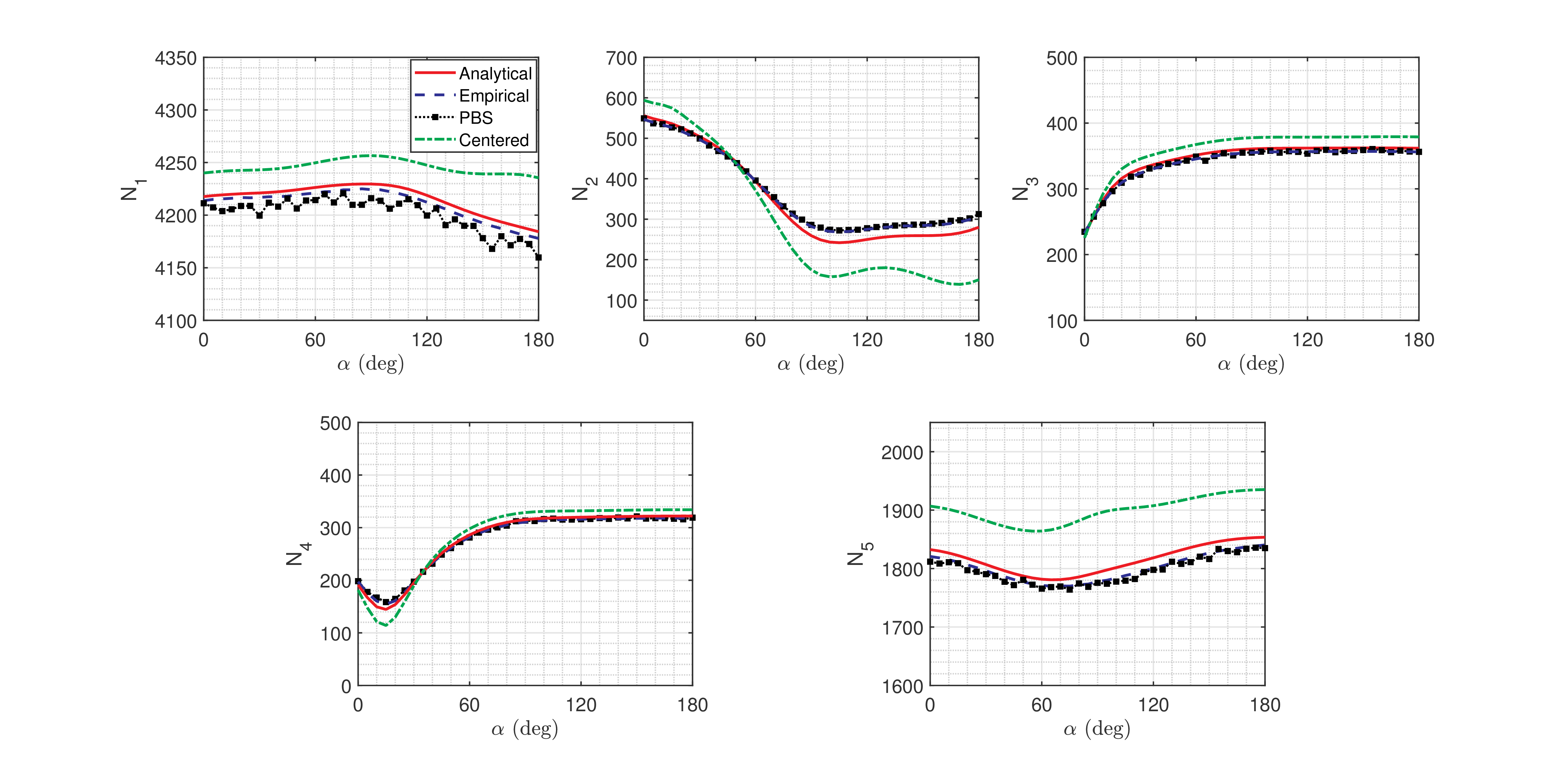}
    \caption{Cumulative expected number of molecules $N_i$, $i$$\,\in\,$$\{1,2,3,4,5\}$ by all five receivers according to topology depicted in Fig.~\ref{fig:topo_si50}. Receiver $\mathcal{R}_2$'s position varies according to the angle $\alpha$, which is the relative angle between the transmitter located in the origin of the coordinate and the center of $\mathcal{R}_2$. Simulation time is $t$$\,=\,$$2\,\mathrm{s}$.  The legend of the figures is defined the same as the ones explained in Fig.~\ref{fig:Same_R}.}
    \label{fig:Si5o}
\end{figure}

 In Fig.~\ref{fig:Si5o} we plot the cumulative number of absorbed molecules by the five receivers at $t$$\,=\,$$2\,\mathrm{s}$. The legend of the figures is the same as the one explained in Fig.~\ref{fig:Same_R}. Each subplot corresponds to the cumulative number of absorbed particles by one of the receivers. The horizontal axis is based on the angle $\alpha$ in degrees. In all subfigures we can observe a perfect match between the data obtained from the empirical barycenter and PBS results. This observation once again approves the idea behind the concept of negative point source and its ideal position, which is the barycenter. Furthermore, the red curve captures the dynamic of the empirical data and shows a very good estimation of the cumulative number of absorbed molecules. There are slight differences for receivers' observations, but given the value of the mismatch it can be considered tolerable in most practical cases. Moreover, the green dash-dot line shows a considerable deviation with respect to the PBS. This observation demonstrates the importance of barycenter analysis and consideration otherwise it is not recommended at all to neglect the barycenters and substitute them with the center of the receivers.
\section{Conclusion}\label{sec:conclusion}
The absorption effect of fully absorbing (FA) receivers can be deduced in terms of a negative point source when there are multiple FA receivers in the channel and the best position to locate this negative point source is the barycenter. Barycenter is defined as the spatial average of the particles that are absorbed by the corresponding receiver. Localizing the barycenter is complex in diffusive MC systems because the presence of receivers brings interaction among them due to the FA characteristic. In this paper we focused on modeling the position of the barycenters in a diffusive molecular communication with multiple FA receivers. The knowledge of the barycenters allows us to solve the system of equations that describes the cumulative number of absorbed molecules. First, we derived an expression that finds the barycenter in the Single Input Single Output system. Then we impose the superposition principle and inspired by Newton's universal gravitational law we equalize the contribution of the sources (negative and positive ones) and verify the resultant model with data obtained from particle-based simulations. The simulation scenarios were chosen to be the most challenging ones, meaning that when the source and the receivers are very close but we obtained very promising results that confirm our modeling approach. We believe that the result of this paper will allow the research community to investigate the systems with multiple FA receivers. Moreover, they can also check whether it is possible to skip even the computation of the barycenter given the simulation parameters based on the analysis of the $\gamma$ discussed in the paper. If the distance between the receivers and the transmitter is not extremely close and sufficient time has passed, one can assume that the barycenters are located in the centers of the receivers. The sufficient time and non-extreme distance can be investigated based on the $\gamma$ provided in the paper.
\appendices
\section{Proof of limits}\label{Ap:proof1}
Let us split~\eqref{eq:Y} into a summation as $A$$\,+\,$$B$ while
\begin{equation}
    A = \frac{R^2w}{2\left(wR+D\right)}\frac{\alpha+\beta R}{\alpha^{3/2}}\mathrm{erfc}\left(\frac{k}{2\sqrt{t}}\right) ,
\end{equation}
\begin{equation}
    B = \frac{R^2w}{2}\frac{wk^2-\beta(1-mk)}{\alpha Dmk} \times \exp{\left(mk+m^2t\right)}\mathrm{erfc}\left(m\sqrt{t}+\frac{k}{2\sqrt{t}}\right). \label{eq:B}
\end{equation}
 The limit of $A$ is straightforward
\begin{equation}
    \lim_{w\to \infty}A = \frac{\alpha R+\beta R^2}{2\alpha^{3/2}}\mathrm{erfc}\left(\frac{k}{2\sqrt{t}}\right).
\end{equation}
According to~\cite{cody1993algorithm} the complementary error function is
\begin{equation}\label{eq:erfcx}
    \mathrm{erfc}(x) = e^{-x^2}\mathrm{erfcx}(x),
\end{equation}
where $\mathrm{erfcx}(x)$ is the scaled complementary error function. Substituting the complementary error function according to~\eqref{eq:erfcx} in $B$ leads to
\begin{align}
     \resizebox{0.85\hsize}{!}{$\begin{aligned}
    B &= \frac{R^2w}{2}\frac{wk^2-\beta(1-mk)}{\alpha Dmk} e^{\left(mk+m^2t\right)}e^{(-m^2t-\frac{k^2}{4t}-mk)}
    \mathrm{erfcx}(m\sqrt{t}+\frac{k}{2\sqrt{t}}) ,\\
    &= \frac{R^2w}{2}\frac{wk^2-\beta(1-mk)}{\alpha Dmk}e^{(-\frac{k^2}{4t})}\mathrm{erfcx}(m\sqrt{t}+\frac{k}{2\sqrt{t}}) ,\\
    &= \frac{R^2w}{2}\frac{wk^2-\beta(1-\frac{wR+D}{\sqrt{D}R}k)}{\alpha \sqrt{D}\frac{wR+D}{R}k}\times e^{(-\frac{k^2}{4t})}\mathrm{erfcx}(\frac{wR+D}{\sqrt{D}R}\sqrt{t}+\frac{k}{2\sqrt{t}}) .
    \end{aligned}$}\label{eq:erfcx_sub}
 \end{align}
As the $w\to \infty$ obviously the argument of $\mathrm{erfcx}$ goes to infinite. The following approximation is valid for large argument of the scaled complementary error function 
\begin{equation}
    \mathrm{erfcx}(x) \approx \left(\frac{1}{\sqrt{\pi}x}\right) .
\end{equation}
Hence~\eqref{eq:erfcx_sub} can be written as
\begin{equation}
    B \approx  \frac{R^2w}{2}\frac{wk^2-\beta(1-\frac{wR+D}{\sqrt{D}R}k)}{\alpha \sqrt{D}\frac{wR+D}{R}k}\frac{e^{(-\frac{k^2}{4t})}}{\sqrt{\pi}(\frac{wR+D}{\sqrt{D}R}\sqrt{t}+\frac{k}{2\sqrt{t}})}\label{B_APRX_F}
 \end{equation}
Finally, taking the limit of~\eqref{B_APRX_F}, the limit of $A+B$ can be written as
\begin{equation}
    \lim_{w\to \infty}A+B = \frac{\alpha R+\beta R^2}{2\alpha^{3/2}}\mathrm{erfc}\left(\frac{k}{2\sqrt{t}}\right)+ \frac{R^{2}e^{(-\frac{k^2}{4t})}}{2\sqrt{\pi}}\left(\frac{k\sqrt{D}+\beta}{\alpha \sqrt{Dt}}\right) .\label{hat:A+B}
\end{equation}
\ifCLASSOPTIONcaptionsoff
  \newpage
\fi
\bibliographystyle{IEEEtran}
\bibliography{Bibliography}

\begin{thebibliography}{10}
\providecommand{\url}[1]{#1}
\csname url@samestyle\endcsname
\providecommand{\newblock}{\relax}
\providecommand{\bibinfo}[2]{#2}
\providecommand{\BIBentrySTDinterwordspacing}{\spaceskip=0pt\relax}
\providecommand{\BIBentryALTinterwordstretchfactor}{4}
\providecommand{\BIBentryALTinterwordspacing}{\spaceskip=\fontdimen2\font plus
\BIBentryALTinterwordstretchfactor\fontdimen3\font minus
  \fontdimen4\font\relax}
\providecommand{\BIBforeignlanguage}[2]{{%
\expandafter\ifx\csname l@#1\endcsname\relax
\typeout{** WARNING: IEEEtran.bst: No hyphenation pattern has been}%
\typeout{** loaded for the language `#1'. Using the pattern for}%
\typeout{** the default language instead.}%
\else
\language=\csname l@#1\endcsname
\fi
#2}}
\providecommand{\BIBdecl}{\relax}
\BIBdecl

\bibitem{bi2021survey}
D.~Bi \emph{et~al.}, ``A survey of molecular communication in cell biology:
  Establishing a new hierarchy for interdisciplinary applications,'' \emph{IEEE
  Commun. Surv. Tutor.}, vol.~23, no.~3, pp. 1494--1545, 2021.

\bibitem{akyildiz2019information}
I.~F. Akyildiz \emph{et~al.}, ``An information theoretic framework to analyze
  molecular communication systems based on statistical mechanics,'' \emph{Proc.
  IEEE}, vol. 107, no.~7, pp. 1230--1255, 2019.

\bibitem{soldner2020survey}
C.~A. S{\"o}ldner \emph{et~al.}, ``A survey of biological building blocks for
  synthetic molecular communication systems,'' \emph{IEEE Commun. Surv.
  Tutor.}, vol.~22, no.~4, pp. 2765--2800, 2020.

\bibitem{barros2021molecular}
M.~T. Barros \emph{et~al.}, ``Molecular communications in viral infections
  research: Modeling, experimental data, and future directions,'' \emph{IEEE
  Trans. Mol. Biol.}, vol.~7, no.~3, pp. 121--141, 2021.

\bibitem{yang2020comprehensive}
K.~Yang \emph{et~al.}, ``A comprehensive survey on hybrid communication in
  context of molecular communication and terahertz communication for
  body-centric nanonetworks,'' \emph{IEEE Trans. Mol. Biol.}, vol.~6, no.~2,
  pp. 107--133, 2020.

\bibitem{chude2017molecular}
U.~A. Chude-Okonkwo \emph{et~al.}, ``Molecular communication and nanonetwork
  for targeted drug delivery: A survey,'' \emph{IEEE Commun. Surv. Tutor.},
  vol.~19, no.~4, pp. 3046--3096, 2017.

\bibitem{cao2019diffusive}
T.~N. Cao \emph{et~al.}, ``Diffusive mobile mc for controlled-release drug
  delivery with absorbing receiver,'' in \emph{ICC 2019-2019 IEEE Int. Conf.
  Commun. (ICC)}.\hskip 1em plus 0.5em minus 0.4em\relax IEEE, 2019, pp. 1--7.

\bibitem{guo2020vertical}
W.~Guo \emph{et~al.}, ``Vertical underwater molecular communications via
  buoyancy: Gaussian velocity distribution of signal,'' in \emph{ICC 2020-2020
  IEEE Int. Conf. Commun. (ICC)}.\hskip 1em plus 0.5em minus 0.4em\relax IEEE,
  2020, pp. 1--6.

\bibitem{jamali2019channel}
V.~Jamali \emph{et~al.}, ``Channel modeling for diffusive molecular
  communication—a tutorial review,'' \emph{Proc. IEEE}, vol. 107, no.~7, pp.
  1256--1301, 2019.

\bibitem{koo2016molecular}
B.-H. Koo \emph{et~al.}, ``Molecular mimo: From theory to prototype,''
  \emph{IEEE J. Sel. Areas Commun.}, vol.~34, no.~3, pp. 600--614, 2016.

\bibitem{bao2019channel}
X.~Bao \emph{et~al.}, ``Channel modeling of molecular communication via
  diffusion with multiple absorbing receivers,'' \emph{IEEE Wireless Commun.
  Lett.}, vol.~8, no.~3, pp. 809--812, 2019.

\bibitem{sabu2022channel}
N.~V. Sabu \emph{et~al.}, ``Channel characterization and performance of a 3-d
  molecular communication system with multiple fully-absorbing receivers,''
  \emph{IEEE Trans Commun.}, 2022.

\bibitem{sabu20203}
------, ``3-d diffusive molecular communication with two fully-absorbing
  receivers: Hitting probability and performance analysis,'' \emph{IEEE Trans.
  Mol. Biol.}, vol.~6, no.~3, pp. 244--249, 2020.

\bibitem{jia2022capacity}
Z.~Jia \emph{et~al.}, ``Capacity analysis of diffusive molecular communication
  system with an interfering receiver,'' in \emph{2022 Int. Conf. on Networking
  and Network Applications (NaNA)}.\hskip 1em plus 0.5em minus 0.4em\relax
  IEEE, 2022, pp. 23--28.

\bibitem{Fardad}
M.~Ferrari \emph{et~al.}, ``Channel characterization of diffusion-based
  molecular communication with multiple fully-absorbing receivers,'' \emph{IEEE
  Trans Commun.}, vol.~70, no.~5, pp. 3006--3019, 2022.

\bibitem{redner2001guide}
S.~Redner, \emph{A guide to first-passage processes}.\hskip 1em plus 0.5em
  minus 0.4em\relax Cambridge university press, 2001.

\bibitem{yilmaz2014three}
H.~B. Yilmaz \emph{et~al.}, ``Three-dimensional channel characteristics for
  molecular communications with an absorbing receiver,'' \emph{IEEE Commun.
  Lett.}, vol.~18, no.~6, pp. 929--932, 2014.

\bibitem{saeed2021analytical}
M.~Saeed \emph{et~al.}, ``An analytical propagation model for diffusion-based
  molecular communication systems,'' \emph{IEEE Trans. Mol. Biol.}, 2021.

\bibitem{dinc2019effective}
F.~Dinc \emph{et~al.}, ``The effective geometry monte carlo algorithm:
  Applications to molecular communication,'' \emph{Physics Letters A}, vol.
  383, no.~22, pp. 2594--2603, 2019.

\bibitem{newton1968mathematical}
I.~Newton, I.~B. Cohen, and A.~Motte, \emph{The Mathematical Principles of
  Natural Philosophy... Translated... by Andrew Motte, Etc.[A Facsimile, with
  an Introduction by I. Bernard Cohen, of the Edition of 1729.].}\hskip 1em
  plus 0.5em minus 0.4em\relax Dawsons of Pall Mall, 1968.

\bibitem{Fardad_Asym}
F.~Vakilipoor \emph{et~al.}, ``Asymptotic mimo channel model for diffusive mc
  with fully-absorbing receivers,'' \emph{IEEE Wireless Commun. Lett.},
  vol.~11, no.~8, pp. 1634--1638, 2022.

\bibitem{cody1993algorithm}
W.~J. Cody, ``Algorithm 715: Specfun--a portable fortran package of special
  function routines and test drivers,'' \emph{ACM Trans Math Softw.}, vol.~19,
  no.~1, pp. 22--30, 1993.

\end{thebibliography}

\end{document}